\documentstyle[preprint,aps,tighten]{revtex}
\newcommand{\half}{\frac{1}{2}}
\def\dirac#1{\setbox0=\hbox{$#1$}\rlap{\hbox to \wd0{$\hss\mkern1mu/\hss$}}
\box0 }
\def\adag{a^{\dagger}}
\def\bdag{b^{\dagger}}
\def\q-1{q^{-1}}
\begin{document}
\draft
\title{ Deformed Heisenberg algebra: origin of q-calculus
}
\author{P. Narayana Swamy$^a$}
\address{Professor Emeritus, Department of Physics,Southern
Illinois University, Edwardsville, IL 62026-1654}
 \maketitle
\begin{abstract}

The intimate connection between  q-deformed Heisenberg uncertainty
relation  and the Jackson derivative based on  q-basic numbers has
been noted in the literature. The purpose of this work is  to
 establish  this connection  in a clear and self-consistent
formulation and to show explicitly how the Jackson derivative
arises naturally. We utilize a holomorphic representation to
arrive at the correct algebra to describe q-deformed bosons. We
investigate the algebra of q-fermions and point out how different
it is from the theory of q-bosons. We show that the holomorphic
representation for q-fermions is indeed feasible in the framework
of the theory of generalized fermions. We also examine several
different q-algebras in the context of  the modified Heisenberg
equation of motion.

\end{abstract}
\vspace{2in} $^a$ Electronic address: pswamy@siue.edu\\
March 2003
 \noindent 

\section{Introduction}

Representations of the q-deformed Heisenberg algebra have received
much attention in the literature \cite{Hellstrom} and its
connection with the q-calculus based on  basic numbers have been
investigated by many authors
\cite{Fink,Hebecker,Schwenk,Floratos,Cerchiai}.  However, the
origin of q-calculus has not been established satisfactorily in a
self-consistent formulation.
 Past investigation on this subject has not fully clarified
  what the assumptions are and
 what consequences follow from them.
 In particular, it has not been clearly established that the q-modified
 Heisenberg uncertainty relation directly leads to the q-algebra of
 bosons and to the Jackson Derivative (JD)\cite{jack}.
  Moreover, only  a few of the authors employ the formulation
which is symmetric under $q \longleftrightarrow \q-1$. The
consequences of q-deformed Heisenberg equation of motion, again,
have not been established in a comprehensive  manner. In the
context of isomorphic or holomorphic representations of
q-algebras, the role of q-calculus in the theory of q-fermions has
been omitted in the literature. In this work we propose to remedy
this situation.

In Sec.II, We shall develop  \textit{ab initio} a self-consistent
and self-contained formulation to establish the connection between
basic numbers, the algebra of q-oscillators and the JD. We shall
demonstrate that the q-boson algebra is intimately connected with
the q-deformed Heisenberg relation and that the JD  is intimately
linked to the q-boson algebra.  In Sec.III, we investigate the
theory of q-fermions and show that the q-fermion algebra does not
share these connections. We shall  show that the theory of
generalized fermions, extending beyond the exclusion principle,
provides a convenient framework to introduce a holomorphic
representation, even though it has unexpected classical limits.
Sec.IV is devoted to a study of the consequences of the modified
Heisenberg equation of motion. We shall investigate the important
role played by the modified Heisenberg equation of motion in the
algebra of both q-bosons and q-fermions. It will be shown that the
formulation of q-fermions arises more naturally from the theory of
generalized fermions which do  not obey the familiar exclusion
principle. Sec.V provides a brief summary.

 First we begin with the dilatation operator
\begin{equation}\label{1}
\theta \equiv x \frac{\partial}{\partial x} \, ,
\end{equation}
which evidently satisfies the properties
\begin{equation}\label{2}
    \theta x = x (\theta +1), \quad \theta^r x = x (\theta
    +1)^r,\quad
    f(\theta) x = x f(\theta +1)\, ,
\end{equation}
where $r$ is any number and $f$ is an arbitrary polynomial. It is
easy to show that such properties extend to any monomial of a real
variable $q$, thus
\begin{equation}\label{3}
    q^{\theta} x= x q^{\theta +1}\, .
\end{equation}
This can be further extended to the basic numbers (bracket
numbers) so that
\begin{equation}\label{4}
    [\theta]\, x = x \, [\theta + 1]\,,
\end{equation}
where the basic numbers are defined by
\begin{equation}\label{5}
    [\alpha]\equiv \frac{q^{\alpha} - q^{-\alpha}}{q- \q-1}\, ,
\end{equation}
in terms of $q$,  an arbitrary real number, $0 < q < \infty$. In
the symmetric formulation one can further restrict it to $0 < q <
1$ or $1 < q < \infty $. This formulation is designed to be
symmetric under $q \longleftrightarrow \q-1$. The JD forms the
basis \cite{Exton} of what is referred to in the literature as
q-calculus and is also intimately connected with the basic numbers
which play a fundamental role. We shall now establish that the JD
and the entire framework of q-calculus has its origin in  the
q-deformed Heisenberg uncertainty relation .

\section{q-Heisenberg algebra and JD: q-bosons}

Let us consider the q-deformed uncertainty principle,  the
q-Heisenberg relation,
\begin{equation}\label{6}
    q \, x p - px = i \hbar \, \Delta \, .
\end{equation}
There are at least two choices for $\Delta$ that would preserve
the correct limit (``classical limit") when $q \rightarrow 1$. The
choice $\Delta = 1$ has been studied by Finkelstein \cite{Fink}.
We shall choose $\Delta = q^{-N}$, where $N$ is to be specified
later,  in order to develop the formulation which is symmetric
under $q \longleftrightarrow\q-1$. We accordingly introduce  an
ansatz or a hypothesis,
\begin{equation}\label{7}
q \, xp - px = i \hbar \, q^{-N} \, .
\end{equation}
For $q \not= 1,$ the momentum  $ p= - i \hbar
\partial_x$ has to be replaced by the generalized operator
  $p= - i \hbar\,  D$,
  and accordingly,
\begin{equation}\label{8}
D  x - q \, x D = q^{-N}\, .
\end{equation}
We seek the solution of Eq.(\ref{8}) for the operator $D$ . We can
show that the solution \cite{Fink} is
\begin{equation}\label{9}
    D = \frac{1}{x}\, \frac{q^N - q^{-N}}{q-\q-1}\,.
\end{equation}
This also renders the deformation (\ref{7}) unique. The basic
number occurring here, defined in Eq.(\ref{5}) forms the basis of
the algebra of q-deformed oscillators. The proof is
straightforward and utilizes the property $x \,[N+1]=[N]\, x$. An
alternative proof which is instructive, consists of first showing
\begin{equation}\label{10}
    [N] \, x \,=\, x \, [N+1] \, = \, x \, q\, [N] + x\,  q^{-N}\, ,
\end{equation}
which leads to
\begin{equation}\label{11}
    [N] \, x - q \, x \, [N] = x \, q^{-N}\, .
\end{equation}
From Eqs.(\ref{8},\ref{11}) the solution, $D = x^{-1}\, [N]$
follows immediately. This is the fundamental connection between
the Jackson derivative and the q-basic number: the q-deformation
of the uncertainty relation thus automatically leads to the JD via
the basic number.  In order to show that Eq.(\ref{9}) is indeed
the same as the familiar generalized derivative in the standard
form, we proceed as follows.

Continuing from Eq.(\ref{3}) we can establish the further
property,
\begin{equation}\label{12}
    \theta \, x^r = r x^r\, ,
\end{equation}
for any number $r$, which can be extended to a monomial, thus
\begin{equation}\label{13}
    \theta^a \,  x^r = r^a  x^r\, .
\end{equation}
Upon using the series
\begin{equation}\label{A1}
    q^{\theta}= 1 + \theta \ln q + \frac{\theta^2}{2 !} (\ln q)^2
    + \cdots \, ,
\end{equation}
we obtain,  for a monomial,
\begin{equation}\label{14}
q^{\theta} x^r \; =\; x^r q^r\;  = \; (q \, x)^r\, .
\end{equation}
This relation immediately generalizes to a polynomial and we
obtain Eq.(\ref{12}) for any polynomial function.  We now make the
observation \cite{Fink,Floratos} that there exists the holomorphic
representation
\begin{equation}\label{15}
   \theta \Longleftrightarrow N,\; \theta = x \partial_x \Longleftrightarrow
   \adag a = N, \quad
   {\rm or  }
    \quad   \theta = x\,  D
    \Longleftrightarrow \adag a = [N]\, ,
\end{equation}
where $N$ is the number operator of q-deformed bosons and $a,
\adag$ denote the annihilation and creation operators of q-boson
oscillators. The property, Eq.(\ref{14}),  is also valid for an
arbitrary polynomial function, where $\theta$ can be either $N$ or
$[N]$. If we choose $\theta=N,$ we derive the property
\begin{equation}\label{16}
    q^N f(x)= f(qx)\, ,
\end{equation}
which can be further  extended to the result
\begin{equation}\label{17}
q^{-N} f(x)= f(\q-1 x)\,.
\end{equation}
From this we may immediately obtain the important result
\begin{equation}\label{18}
    D f(x) = \frac{1}{x} \, \frac{q^N - q^{-N}}{q- \q-1}\; f(x) \; =
    \; \frac{1}{x}\,  \frac{f(qx) - f(\q-1 x)}{q- \q-1}\, ,
\end{equation}
which is recognized as the standard definition \cite{Exton,Alpns}
of the JD in the symmetric formulation. It is clear that the JD
reduces to the ordinary derivative in the limit $q \rightarrow 1$.
Hence we have established the connection between the q-Heisenberg
 relation on one hand and the q-basic number and JD on
the other hand in the symmetric formulation. Consequently,  the
introduction of JD arises naturally from the q-deformed Heisenberg
relation via the q-basic number.

Let us now consider q-bosons and the JD which obeys the relation
(\ref{8}). If we choose the representation $D \Rightarrow a, \, x
\Rightarrow \adag $ as in (\ref{15}), then we immediately obtain
\begin{equation}\label{19}
    a \adag - q \adag a = q^{-N}\, .
\end{equation}
Accordingly, the algebra of creation and annihilation operators of
the q-deformed oscillators can be regarded as arising from a
representation of the coordinate and the JD. Thus in this sense,
the q-deformed algebra of bosons is an immediate consequence of
the deformed Heisenberg algebra.

We conclude this section by recording some  results for later use,
\begin{equation}\label{20}
    q^N a = a \, q^{N-1}, \quad [N]\,  a = a\,  [N-1], \quad [N+1]= q \, [N] +
    q^{-N}\, .
\end{equation}

\section{q-fermions}

We shall now  turn our attention to  q-fermions. The statistical
mechanics and thermodynamics of q-deformed fermions has been
investigated thoroughly \cite{Alpns2} on the basis of q-deformed
algebra.
 Let us specifically consider the basic number as
defined in Eq.(\ref{5}) while the creation and annihilation
operators satisfy \cite{Alpns2} the algebra
\begin{equation}\label{21}
    b \bdag + \q-1 \bdag b = q^{-N}\, .
\end{equation}
This formulation which leads to the determination of many
thermodynamic functions of q-bosons and q-fermions as well as to
interesting predictions, displays a clearly desirable symmetry
among fermions and bosons and moreover the basic number used for
fermions is exactly the same as for bosons.  When we examine the
algebra described by Eqs.(\ref{5}) and (\ref{21}), together with
the Heisenberg relation, we conclude that the modified Heisenberg
uncertainty relation such as in Eq.(\ref{7}), or one of its
variants,  does not provide a representation for the JD (\ref{9})
to produce the q-fermion algebra (\ref{21}). In other words, there
does not exist any useful holomorphic representation for the JD,
 as in Eq.(\ref{15}), such as $x \Leftrightarrow
  b^{\dag}, \, \partial_x \Leftrightarrow b $ that will produce
  the algebra in
Eq.(\ref{21}). On the other hand it should be noted that the JD in
the case of q-fermions still arises from the q-deformed Heisenberg
relation as was shown before. In the theory based on this algebra
of q-fermions, the eigenvalues of the number operator $N$ can take
the values $n=0,1$ only. Thus the theory obeys Pauli exclusion
principle just as in  the case of undeformed fermions.
Furthermore, it can be shown that algebras such as above can be
transformed to the case of undeformed ordinary fermions
\cite{Partha,Chaichian} and thus it would appear that such
algebras may not represent genuine deformations or
generalizations. Such transformations themselves might, however,
be of some interest \cite{pns2}.

There exists another interesting representation  resulting in a
different q-deformed fermion algebra which we shall now
investigate. Let us introduce the definition of  the fermion basic
numbers \cite{Chaichian} by
\begin{equation}\label{22}
    [z]_F= \frac{q^{-z}- (-1)^z q^z } {q + \q-1}\, ,
\end{equation}
which has many useful and interesting properties. In this case we
obtain the result that the equation analogous to Eq.(\ref{8}) for
the boson case,
\begin{equation}\label{23}
    B x + q \, x B = q^{-N}\, ,
\end{equation}
has the solution given by  $ B=x^{-1}\, [N]_F $.   This can be
verified as follows. First we confirm that the properties in
Eqs.(\ref{1}-\ref{3}) and (\ref{12}-\ref{14}) are also valid in
the fermion case when we use the generalized basic numbers,
Eq.(\ref{22}). We need an extension of these basic properties. By
writing $(-q)^N$ as $(e^{i \pi}q)^N$ and using the series form for
the exponential, we derive
\begin{equation}\label{A2}
    (-q)^N \, x=  x (-q)^{N+1}\, .
\end{equation}
It should be stressed that $N$ is an operator and the above
equation is in no conflict with Eq.(\ref{3}).
 Consequently, we indeed have the
holomorphic representation $(1/x) [N]_F \Longleftrightarrow b, \,
x \Longleftrightarrow \bdag$ which leads to
\begin{equation}\label{25}
b \bdag + q \bdag b =
    q^{-N} \, ,
\end{equation}
as can be readily verified.  We can also provide a direct proof
for the solution of Eq.(\ref{23}).
 The explicit proof begins with the observation that $[N]_F \, x = x \,
[N+1]_F$, together with the property $q [N] + [N+1] = q^{-N}$ and
then showing
\begin{equation}\label{26}
    q\,  x \, \frac{1}{x}\,  [N]_F + \frac{1}{x}\, x \, [N+1]_F= q^{-N}\, .
\end{equation}
Eq.(\ref{25}) follows immediately. Some observations are in order
here. The algebra (\ref{25}) which can be regarded as resulting
from the representation stated above, arises from the special
fermion basic number defined in  Eq.(\ref{22}),  which is
precisely the one introduced by Chaichian et al \cite{Chaichian}.
Secondly, it is remarkable that this fermion basic number readily
provides  a consistent holomorphic representation in terms of the
creation and annihilation operators for the q-fermions, hence
providing a fundamental justification for the definition
(\ref{22}) itself. Thirdly,we note that this q-fermion algebra is
not based on the q-Heisenberg relation. The connection with the
q-Heisenberg relation is valid for q-bosons only. In spite of
appearances, the operator $B$ appearing in Eq.(\ref{23}) has no
relation to the JD, which has its connection only to the q-bosons;
it does not have any resemblance to the ordinary derivative in the
classical limit. Finally we observe that
 this leads to a generalization of
fermions \cite{Chaichian} in the sense that this q-deformed
algebra goes beyond the Pauli exclusion principle and the
eigenvalues of $N$ are arbitrary
 and not restricted to the values $0, 1$. It must
be stressed that the fermion basic number introduced above does
not have the expected classical limit, $\lim_{q \rightarrow 1}\,
[N]_F \not= N$. Indeed we find that when $q \rightarrow 1$,
\begin{equation}\label{27}
[N]_F \longrightarrow \, \frac{1- (-1)^N}{q+\q-1}\quad
\longrightarrow \, \half \{ 1 - (-1)^N \} \, ,
\end{equation}
which reduces to unity when $N$ is odd and vanishes when $N$ is
even. The limits in more general cases are described in
ref.\cite{Chaichian}. We reiterate that while the q-boson algebra,
the q-deformed Heisenberg relation and the JD are intimately
connected, the q-calculus  for q-fermions has a different origin
and is not related to the q-Heisenberg relation.

\section{ Heisenberg equation of motion and q-algebra}
The Heisenberg equation of motion (H-equation)
\begin{equation}\label{28}
i \hbar \dot{F} \; = \; i \hbar \frac{dF}{dt} \;= \; [F,H]\;  = \;
FH - HF\, ,
\end{equation}
determines the time evolution of an operator, and hence  the
dynamics, in standard quantum mechanics. In a q-deformed theory,
we expect this to be modified according to a form such as
\begin{equation}\label{29}
    i \hbar \dot{F}= FH - qHF\, ,
\end{equation}
which is not unique since many variations of this form might have
the same classical $q \rightarrow 1 $ limit. If we consider the
boson oscillator Hamiltonian, $H= \half \hbar \omega (\adag a + a
\adag)$ in the above equation, together with the q-boson algebra
$a \adag - q \adag a = 1$ in the nonsymmetric formulation, as in
ref. \cite{Fink} we obtain
\begin{equation}\label{30}
    i \hbar \dot{a}= a H - q H a = \hbar \omega a \,,
\end{equation}
which then yields $a \sim e^{- i \omega t}$ and everything is
consistent. On the other hand, in the formulation symmetric under
$q \longrightarrow \q-1$, we might work alternatively with the
undeformed H-equation,
\begin{equation}\label{31}
i \hbar \dot{F}= [F,H]\, ,
\end{equation}
as a variation of the above, together with the algebra $a \adag -
q \, \adag a = q^{-N}$ and the Hamiltonian of q-bosons, $H= \half
\hbar \omega \, ( \, [N] + [N+1]\, )$. We then obtain in this
case,
\begin{equation}\label{32}
    i\hbar\,  \dot{a}= \hbar \omega \, a \; , \frac{q^N + q^{N-1} } {2}\; e^{-i \omega
    t},\,
\end{equation}
which has the solution
\begin{equation}\label{33}
    a = a_0\,  \frac{q^N + q^{-N}}{2}\,  e^{-i \omega
    t}\,,
\end{equation}
where the constant $a_0 $ is independent of $q$. On the other hand
if we modify the H-equation as in Eq.(\ref{30}), we obtain, for
the same theory of q-bosons,
\begin{equation}\label{34}
 a = a_0 \, \frac{q^{-N-1} +
q^{-N}} {2} \; e^{-i \omega
    t}\, .
\end{equation}
Thus there is no significant difference from (\ref{33}) in the
time dependence or the q-dependence in these variants.  While the
deformed H-equation is thus non-unique, for the sake of
definiteness, we shall henceforward  make an arbitrary choice and
adhere consistently to the following modified H-equation:
\begin{equation}\label{35}
i \hbar \, \dot{F}= FH - \q-1 HF\, .
\end{equation}
First let us consider the system of q-bosons  described by
\begin{equation}\label{36}
    H= \half \hbar \omega \,  (\adag a + a \adag ) = \half \hbar
    \omega \, (\, [N] + [N+1] \,  )\, .
\end{equation}
If we employ the deformed H-equation, Eq.(\ref{35}),  we obtain
\begin{eqnarray}
  i\hbar \, \dot{a} &=& \half \hbar \omega
  \left \{ a\, ([N]+[N+1])- \q-1([N]+[N+1])\, a \right \}\nonumber \\
   &=& \half \hbar \omega \, a \, (q^N + q^{-N})\, .
  \label{37}
\end{eqnarray}
Consequently the time dependence is given by
\begin{equation}\label{38}
 a = a_0 \, \frac{q^{N} +
q^{N-1}} {2}\,  e^{-i \omega
    t},
\end{equation}
where $a_0$ is independent of $q$.

  We shall next consider the
case of q-fermions. The system of ``standard" q-fermions as in
ref. \cite{Alpns2} is described by the Hamiltonian,
\begin{equation}\label{39}
    H= \half \hbar \omega \, (\adag a - a \adag) \; = \; \half \hbar
    \omega \, ([N] - [1-N])\; ,
\end{equation}
which has the ``correct" classical limit $q \rightarrow 1$, $\lim
\, H = \pm \half \hbar \omega , $ corresponding to the eigenvalues
$N= 0, 1 .$ If we employ Eq.(\ref{35}), we obtain in this case,
\begin{eqnarray}
  i\hbar \omega \, \dot{a} &=& \half \hbar \omega\,
  \left \{ a \left ([N] - [1-N] \right ) - \q-1 \left ([N] - [1-N] \right ) a   \right \} \nonumber\\
   &=& \half \hbar \omega \left \{ q^{N-1} + q^{N-2} \right \} \, .
   \label{40}
\end{eqnarray}
Accordingly the time dependence is
\begin{equation}\label{41}
a = a_0 \, \frac{q^{N-1} + q^{N-2}}{2} \, e^{-i \omega
    t},
\end{equation}
in accordance with expectations.

Let us finally  consider the Chaichian formulation\cite{Chaichian}
of generalized q-fermions  which is  described by
\begin{equation}\label{42}
\adag a = [N]_F, \; a \adag = [N+1]_F, \; H= \half \hbar \omega\,
\left ([N]_F - [N+1]_F \right ) \, .
\end{equation}
We note the classical limit, $q \rightarrow 1$, of the Hamiltonian
is  $H \rightarrow H_0 = \half \hbar \omega$ , in contrast to the
case of  the standard version where
\begin{equation}\label{43}
    H_0 = \half \hbar \omega (2 N - 1) = \pm \half \hbar \omega\,
    .
\end{equation}
 Consequently, this system  not only  describes generalized fermions
 which are not constrained by  the exclusion principle, it  also does not
 behave like  the standard system of fermions in the
 classical limit. Upon employing the
H-equation, Eq.(\ref{35}), we obtain the result
\begin{equation}\label{44}
    a= a_0 \, \frac{(-q)^N - (-q)^{N-1}}{2} \, e^{- i \omega t}\, ,
\end{equation}
which is significantly different from (\ref{41}) in terms of the
q-dependence. To see that it is so, we might  examine the
classical limit,  $q \rightarrow 1$, and  obtain $a = a_0 (-1)^N
e^{-i \omega t}$. It looks  similar to the expected classical
limit but the phase alternates between $1$ and $-1$ for the
allowed values $N=0, 1, 2, \cdots \infty$. However, the theory is
self-consistent.  Hence we conclude that in each of the above
three cases, the deformed H-equation leads to acceptable time
evolution of the creation and annihilation operators consistent
with the corresponding q-algebra and the appropriate definition of
the basic numbers. In particular, the time dependence of the
annihilation and creation operators depend implicitly on $q$ and
$N$, where the latter quantity in turn depends on q.

\section{Summary}

In this work we have shown explicitly that the q-deformed
Heisenberg uncertainty relation, together with the q-deformed
algebra of boson oscillators leads naturally to the Jackson
Derivative. We  have established this intimate connection in a
self- contained and comprehensive  formulation, symmetric under $q
\longleftrightarrow \q-1$. We also showed that the q-deformed
boson algebra itself can be regarded as arising from a holomorphic
representation for the creation and annihilation operators. We
have thus established the basis of q-calculus as a direct
consequence of the basic numbers.

 Upon
examining the algebra of q-fermions we concluded that it has no
direct link with the q-deformed Heisenberg relation. For the
q-fermions, the fermion basic numbers are connected to the
q-fermion algebra by a holomorphic representation in a
self-consistently formulated theory of generalized fermions which
go beyond the exclusion principle.
 Finally we
have examined the Heisenberg equation of motion, deformed for the
purpose of describing q-deformed systems. We have studied three
different algebras in this context and we have shown that the
q-bosons as well as the generalized fermions do possess desirable
properties.

\acknowledgments

I thank A. Lavagno of Politecnico di Torino, Italy,  for fruitful
discussions on the subject of the Jackson derivative and q-bosons
and q-fermions.

\end{document}